# Development of an X-band Photoinjector at SLAC*

**A. E. Vlieks, G. Caryotakis, R. Loewen, D. Martin, A. Menegat**
*SLAC, 2575 Sand Hill Rd, Menlo Park, CA 94025, USA*
**E. Landahl, C. DeStefano, B. Pelletier, and N.C. Luhmann, Jr.**
*3001 Engineering III, Dept. of Applied-Science*
*Davis, CA 95616, USA*

*Abstract*
As part of a National Cancer Institute contract to develop a compact source of monoenergetic X-rays via Compton backscattering, we have completed the design and construction of a 5.5 cell Photoinjector operating at 11.424 GHz. Successful completion of this project will result in the capability of generating a monoenergetic X-ray beam, continuously tunable from 20 - 85 KeV. The immediate goal is the development of a Photoinjector producing 7 MeV, 0.5nC, sub-picosecond electron bunches with normalized RMS emittances of ≃.1 pi-mm-mR at repetition rates up to 60 Hz.. This beam will then be further accelerated to 60 MeV using a 1.05 m accelerating structure. This Photoinjector is somewhat different than the traditional 1.5 cell design both because of the number of cells and the symmetrically fed input coupler cell. Its operating frequency is also unique. Since the cathode is non-removable, cold-test tuning was somewhat more difficult than in other designs. We will present results of "bead-drop" measurements used in tuning this structure.

Initial beam measurements are currently in progress and results will be presented as well as results of RF conditioning to high gradients at X-band. Details of the RF system, emittance-compensating solenoid, and cathode laser system as well as PARMELA simulations will also be presented.

## 1. Introduction

We have recently completed the design and construction of the main components of an X-band photoinjector and are currently in the initial stage of testing. In this paper we will describe the design and initial testing of the RF gun. The essential components of the Photoinjector are shown in Figure 1. The 0.6 T Emittance compensation Solenoid consists of a pair of identical coils positioned axially with respect to each other. The coils are oppositely wound so that the field is zero in the center of the solenoid. The RF gun is positioned in the Solenoid with the cathode at this field null. A Laser Mirror chamber permits a UV laser beam to strike the cathode along a nearly axial path. The generated electron beam is further accelerated by a 1.05 m accelerating structure. At the exit of the accelerator three quadrupoles are positioned to focus the electrons to a narrow spot size in the Interaction chamber. A Spectrometer, not shown, is used to divert the beam away from its axial path into a beam dump.

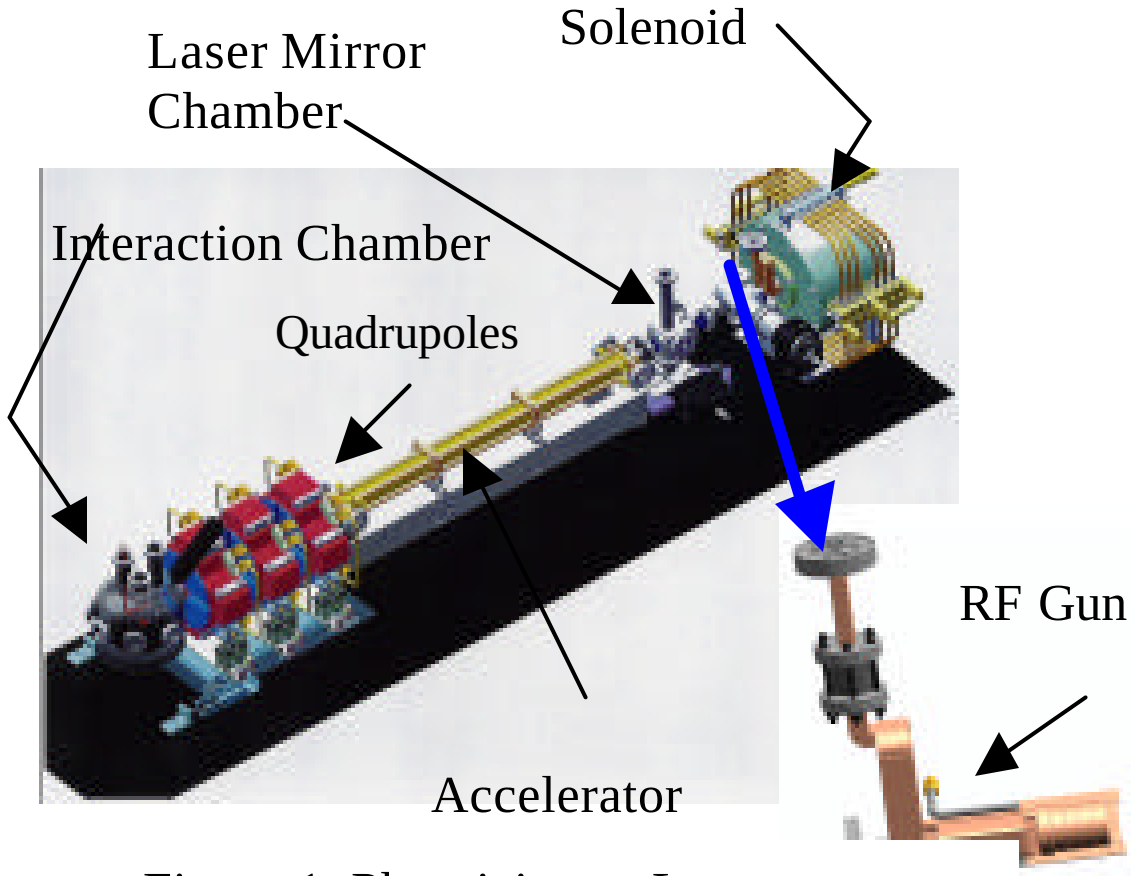

Figure 1. Photoinjector Layout

This RF gun is unique in its design because of the number of cells used and its high operating surface field gradients. The basic gun parameters are listed in table 1. The RF power for the RF gun and accelerator will be derived from two 60 MW X-band Klystrons driven by a common frequency generator. This will permit independent phase and amplitude control of both the accelerator and Gun RF.

| | |
|---|---|
| Number of Cells | 5.5 |
| Peak Surface Gradient/Power | 200 MV/m @ 16 MW |
| RF Filling Time | 65 ns |
| Cathode Material | Copper |
| RF Pulse length | 200 ns |

Table 1. RF Gun Parameters

## 2. Simulations

The initial gun design was obtained using SUPERFISH[1]. It was used to identify the 6 modes and establish the pi-mode at the correct frequency. (11.424 GHz.) It was also used to determine the Q-value, the R/Q and the frequency separation of the pi-mode from its nearest neighbor. HFSS[2] and MAFIA[3] were used to design the symmetrically fed coupler cell and establish the desired external Q of the gun. PARMELA[4] was used to optimize the beam focusing and emittance as well as establish the approximate location of the accelerator, quadrupoles and interaction chamber. The required design strength of the solenoid and quadrupoles were also determined using PARMELA. Figure 2 shows the electron beam profile as it is accelerated through the gun and a .75 m accelerator structure and finally focused at the interaction point. This particular simulation was performed assuming a well formed "beer barrel" shaped laser spot at the cathode. The resulting beam spot size at the interaction point was approximately 30 microns with a normalized emittance of 0.75 pi-mm-mrad.

*Work supported by Department of Energy contract DE-AC-76SF00515
and National Cancer Institute Contract Number N01-CO-97113

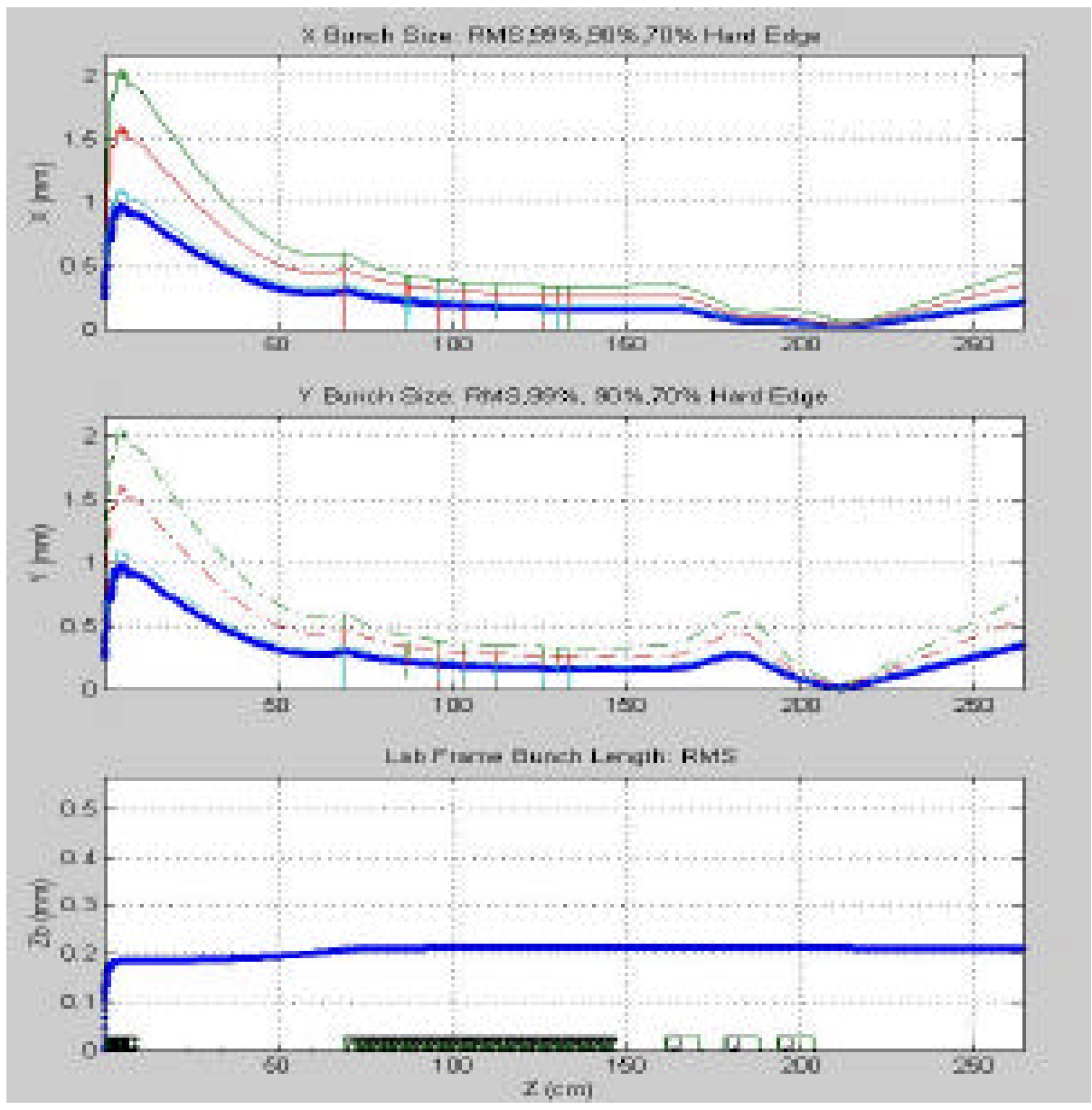

Figure 2. Beam profile along Injector axis.

For a Gaussian-shaped laser spot, the electron beam diameter and emittance increases to 40 microns and 1.1 pi-mm-mrad. For both cases, the laser spot at the cathode had transverse dimensions of 1 mm and a temporal (longitudinal) dimension of 800 fs.

## 3. Cell Tuning

The tuning of the final RF gun proceeded through several steps. Initially a prototype was built using the dimensions from the simulations. The 2b dimension was deliberately undersized to permit frequency tuning by removing material. This was done on a cell-to-cell basis using pretuned end caps. The prototype was then assembled and the coupler cell (cell 6) was retuned to obtain the design external Q value and frequency. The changes required in the final 2b dimension were small enough that an etching procedure was used. A bead-pull measurement was then used to verify that the peak fields in each cell were near identical. The cells were then diffusion bonded and remeasured. The only change due to bonding was to lower the pi mode frequency by 2 MHz. The resulting cavity dimensions were used for the actual RF gun cells.

Tuning posts were incorporated into each cell (4 for the first 5 cells and 2 for the coupler cell).

The individual cell frequencies were measured but no attempt was made to tune the individual cells since they were already within 2 MHz. of their design values. The cells were diffusion bonded and the coupler cell irises were machined to its design value. The peak field flatness was measured using a "bead drop" scheme. In this method a small metal cylinder was lowered into the gun using a precision stepper motor. The frequency shift was measured by a HP8510 network analyzer. Operation of the stepper motor and sampling of the Network Analyzer was performed using a LabVIEW program. The effect of the mass of the bead and the diameter of the supporting string were investigated to find a combination, which gave reasonable and repeatable results. The final choice of string was .02mm nylon suture thread. A comparison of "bead-pull" and "bead-drop" measurements was made on the cold-test prototype (which had an axial hole in the cathode). This verified that the two measurement techniques gave the same results. Figure 3 shows the final field profile after tuning and Figure 4 shows the total measured frequency spectrum

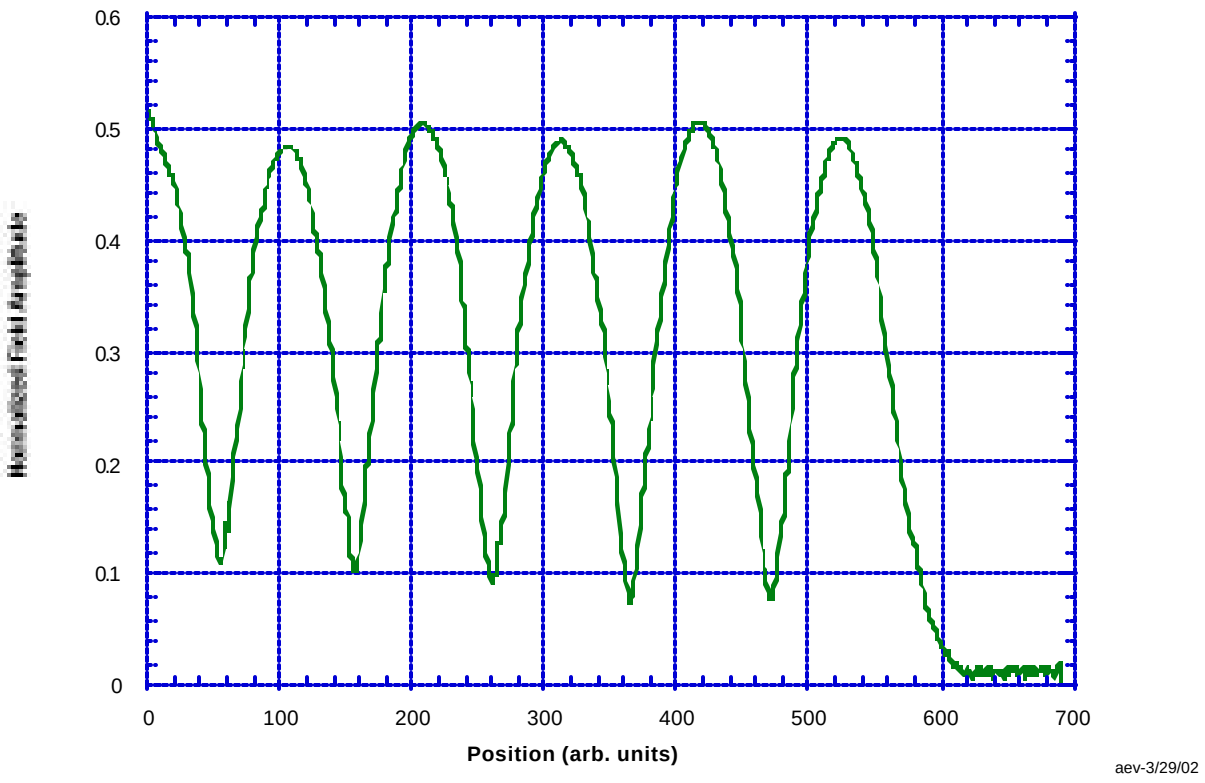

Figure 3. Final post-braze Field profile, measured using "bead-drop" technique

## 4. Gun Preparation

Since the operating peak surface gradient of the RF Gun is rather high, 200MV/m, careful attention to surface cleanliness is necessary. After cold test measurements were completed, the gun was vacuum fired at 750 C. for 8 hours to remove hydrogen and surface contamination. An RF window assembly was then connected to the input waveguide, a vac-ion pump was installed at one port of a Tee at the Gun output beamline and a Gate valve was connected to the other port. See Figure 1. The total assembly was then baked out in a vacuum oven at 130 degrees for 24 hours to remove water vapor. While under vacuum, the valve was closed so that the assembly could be installed in the Solenoid magnet while remaining under vacuum. In this way, the gun was high power conditioned without breaking vacuum.

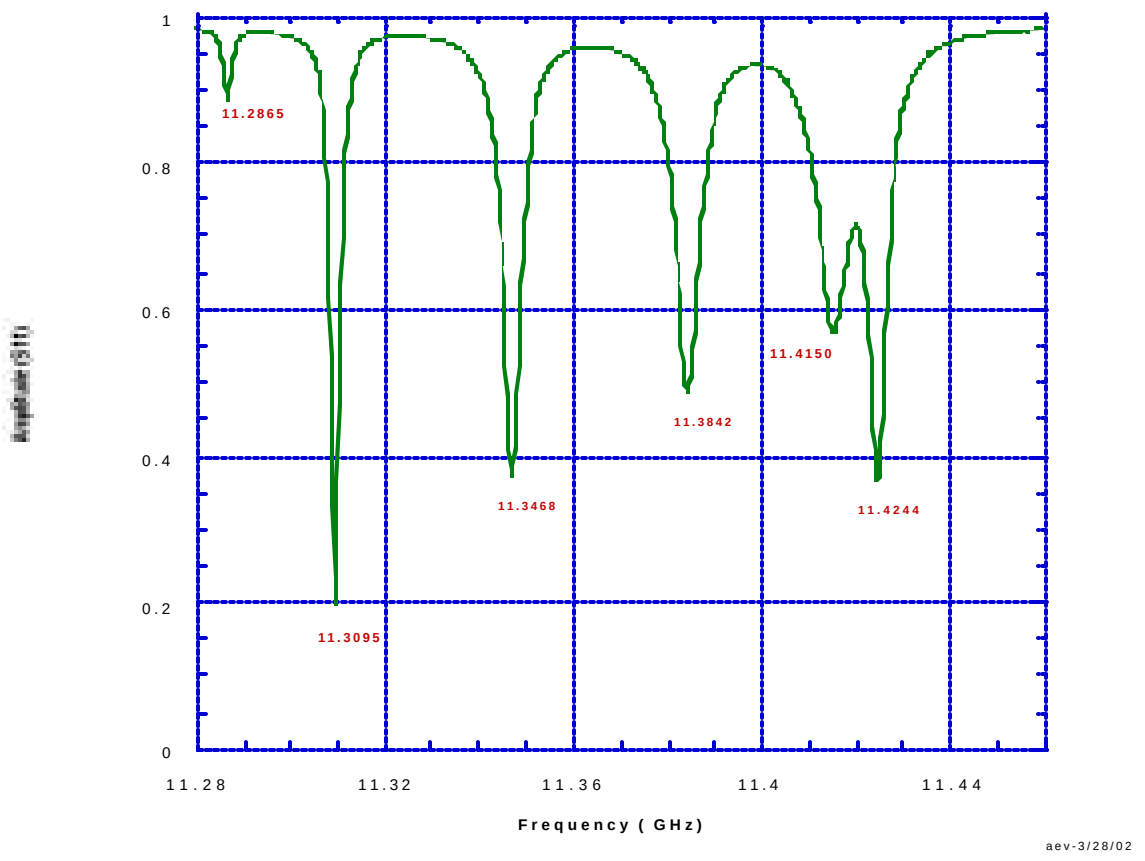

Figure 4. Frequency profile of RF Gun

## 5. High Power Conditioning

After briefly running at 5 MW with a wide, 400ns, RF pulse to study the pulse shapes, the gun was run up to 24 MW with a 75-80 ns pulse. At this power level it was run for 46 minutes without breakdown events. The peak surface gradient at this power level was 168 MV/m. The power was then lowered and the pulse width widened to 100 ns. The power was then increased to 20 MW or 175 MV/m. Again lowering the power, the pulse width was widened to 150 ns. The power was gradually raised to 16-16.5 MW or 180 MV/m. Finally the pulse width was widened to its design value of 200 ns and run up to 15 MW. This corresponds to 185MV/m. Little faulting was observed Final processing will continue at a later date. Typical pulse shapes of the incident and reflected RF signals are shown in Figures 5 and 6. As can be seen from Figure 5, the pulse is flat for the first 200 ns and then suffers from significant amplitude variations. This is caused by the reflected signal returning to the Klystron, and disturbing the subsequent Klystron output power. The round trip distance to the Klystron is almost exactly 200 ns. For our purposes this will not present a problem since we require only a 200 ns pulse. For added flexibility however, additional waveguide length will be added in the near future.

In Figure 6, the typical general form of a reflected RF signal from a standing wave cavity can be seen. The initial and final spikes are due to the filling and emptying of the cavity while the intermediate region shows the approach to equilibrium of an over-coupled cavity.

The spike in the middle is due to the perturbation caused by the forward power pulse shape. It is interesting to note that the initial and final spikes are actually double peaked. This is due to the shock excitation of the neighboring resonance of the pi mode. This was verified by performing a Fourier transformation to the frequency spectrum of Figure 4 with a realistic input pulse shape. The resulting time domain pulse shape is virtually identical to Figure 6. (in the absence of the forward power excursion at the 200 ns region).

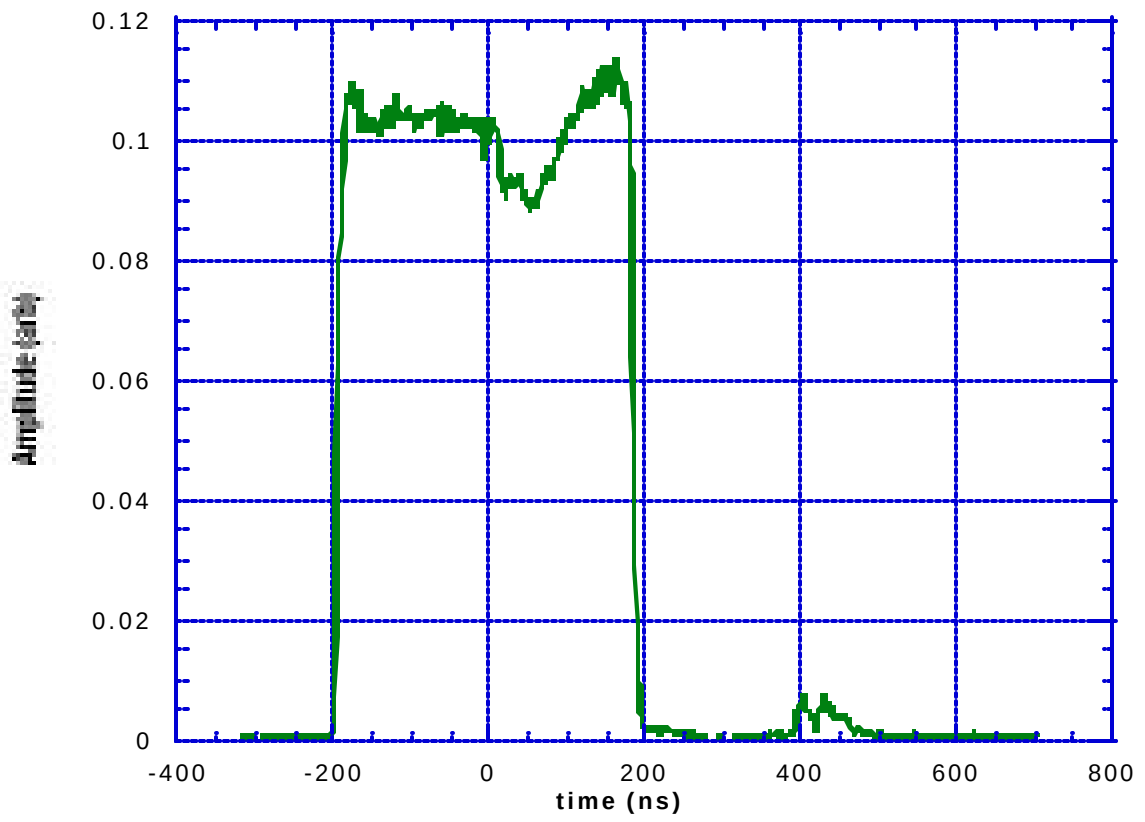

Figure 5. Gun incident RF Power

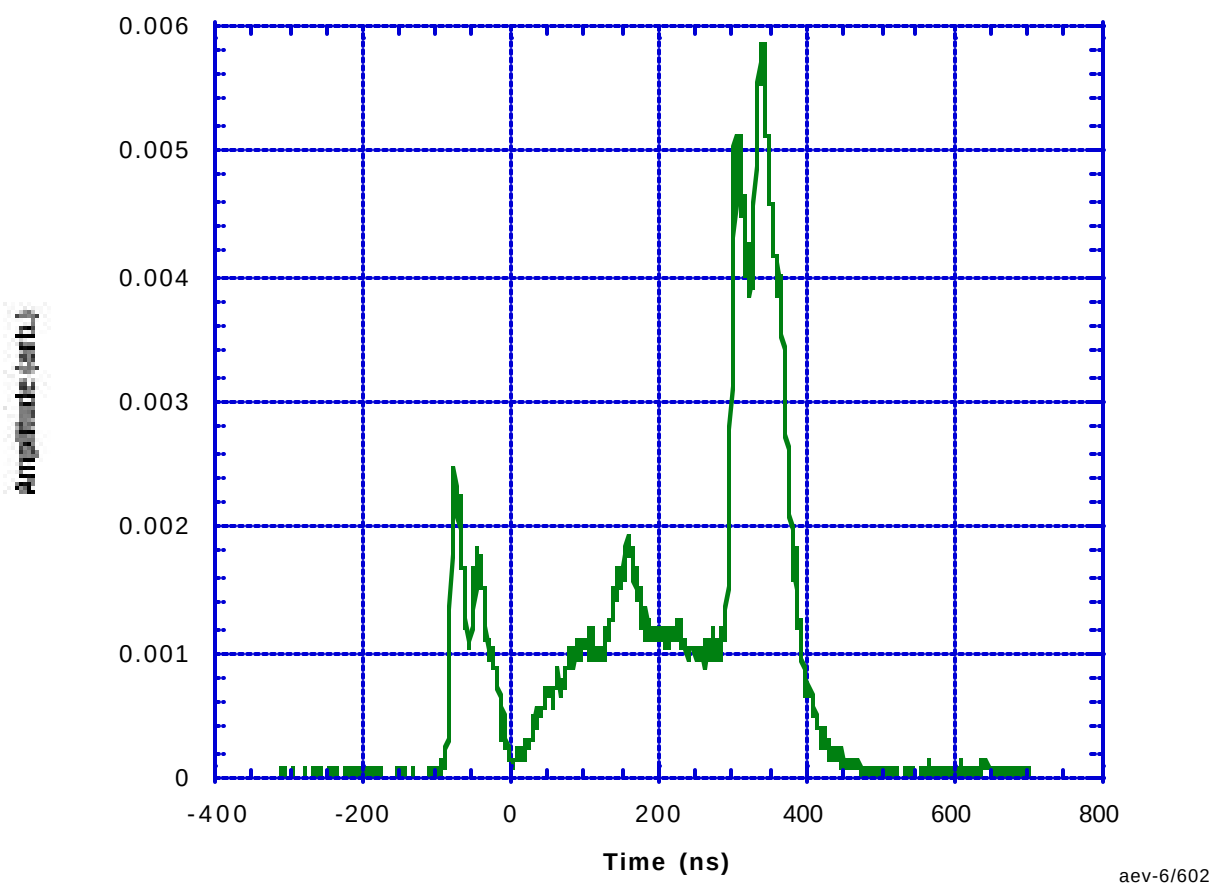

Figure 6. Reflected RF Power from Gun

## 6. References

[1] James Billen and Lloyd M. Young, *POISSON, SUPERFISH reference manual*, LA-UR-96-1834.

[2] Agilent, High-Frequency Structure Simulator v5.6

[3] R.Klatt, F.Krawczyk, W.R.Novender, C.Palm, T.Weiland, "MAFIA- A 3-D Electromagnetic CAD system for Magnet, RF Structures and Transient Wake-Field Calculations", *Reports at the 1986* ***Linear Accelerator Conf***, Stanford, USA 6/2-6/1986

[4] Lloyd Young, PARMELA *reference manual*, LA-UR96-1835, January 8, 2000,. The version of PARMELA used in this work is a modified version due to E. Colby, at SLAC